\documentclass[twocolumn]{revtex4}
\usepackage{epsfig}
\usepackage{graphicx}
\usepackage{amssymb}
\usepackage{amsmath}
\usepackage{amsfonts}
\usepackage{mathrsfs}

\newcommand{\R}{\mathbb{R}}
\newcommand{\C}{\mathbb{C}}

\newcommand{\cH}{\mathcal{H}}

\newcommand{\cP}{\mathcal{P}}

\newcommand{\cT}{\mathcal{T}}

\newcommand{\be}{\begin{equation}}
\newcommand{\ee}{\end{equation}}
\newcommand{\bea}{\begin{eqnarray}}
\newcommand{\eea}{\end{eqnarray}}
\newcommand{\nn}{\nonumber}
\newcommand{\kt}{\rangle}
\newcommand{\br}{\langle}

\newcommand{\ed}{\end{document}}
\newcommand{\bbr}{\br\!\br}
\newcommand{\kkt}{\kt\!\kt}
\newcommand{\pbr}{\prec}
\newcommand{\pkt}{\succ}

\newcommand{\bi}{\begin{itemize}}
\newcommand{\ei}{\end{itemize}}

\newcommand{\bce}{\begin{center}}
\newcommand{\ece}{\end{center}}

\newcommand{\sC}{\mathscr{C}}

\newcommand{\sH}{\mathscr{H}}

\newcommand{\sS}{\mathscr{S}}
\newcommand{\sT}{\mathscr{T}}

\newcommand{\tad}{\hat{\tilde a}^{\!\mbox{$\tilde\dagger$}}}

\begin{document}

\title{Imaginary-Scaling versus Indefinite-Metric Quantization of\\ the Pais-Uhlenbeck Oscillator}

\author{Ali Mostafazadeh}
\address{Department of Mathematics, Ko\c{c} University, 34450 Sar{\i}yer,
Istanbul, Turkey\\ amostafazadeh@ku.edu.tr}

\begin{abstract}
Using the Pais-Uhlenbeck Oscillator as a toy model, we outline a consistent alternative to the indefinite-metric quantization scheme that does not violate unitarity. We describe the basic mathematical structure of this method by giving an explicit construction of the Hilbert space of state vectors and the corresponding creation and annihilation operators. The latter satisfy the usual bosonic commutation relation and differ from those of the indefinite-metric theories by a sign in the definition of the creation operator. This change of sign achieves a definitization of the indefinite-metric that gives life to the ghost states without changing their contribution to the energy spectrum.
\vspace{2mm}\\
\hspace{10cm}{Pacs numbers: 03.65.Ca, 04.60.-m, 11.10.Ef}
\vspace{2mm}
\end{abstract}

\maketitle

\noindent {\em{\bf 1.~Introduction:}}~The recent renewal of interest in higher derivative theories of gravity \cite{gravity} has provided incentive for reconsidering the old problem of quantizing the classical Pais-Uhlenbeck (PU) Oscillator \cite{PU}. Recently the authors of \cite{be-ma} have proposed a quantization scheme involving non-Hermitian $\cP\cT$-symmetric Hamiltonians \cite{pt,ph,ps-2010} that yields a stable and unitary quantum system for the non-degenerate forth-order PU oscillator. A critical assessment and an alternative quantization of the PU oscillator are given in \cite{smilga-sigma,pla-2010c}. In the present paper we explore the quantum analog of the imaginary scaling trick of  \cite{be-ma} and use its mathematical properties to  develop a consistent ghost-free alternative to the indefinite-metric quantization of the PU oscillator. Our method enjoys general applicability and does not involve complex classical quantities or non-Hermitian Hamiltonian operators.

The forth-order classical PU oscillator is the dynamical system defined by the equation of motion:
    \be
    z^{(4)}+\alpha\,z^{(2)}+\beta\,z=0,
    \label{eq-mo}
    \ee
where $z^{(k)}$ denotes the $k$-th derivative of the real dynamical variable $z$, $\alpha:=\omega_1^2+\omega_2^2$, $\beta:=\omega_1^2\omega_2^2$,
and $\omega_1$ and $\omega_2$ are positive real numbers. The solution of (\ref{eq-mo}) is a linear combination of $e^{\pm i\omega_1t}$ and $te^{\pm i\omega_1t}$ for the degenerate case where $\omega_1=\omega_2$, and $e^{\pm i\omega_1t}$ and $e^{\pm i\omega_2t}$ for the non-degenerate case where $\omega_1\neq\omega_2$.

Eq.~(\ref{eq-mo}) may be obtained from either the higher-derivative Lagrangian: $L=\frac{1}{2}(\ddot z^2-\alpha\,\dot z^2+\beta\,z)$ or the Lagrangian: $L=\frac{1}{2}\,(\dot x^2-\alpha\,x^2+\beta\,z^2)+\lambda(\dot z-x)$ that involves a pair of real dynamical variables $x$ and $z$ and a Lagrange multiplier $\lambda$. The latter enforces the constraint $x=\dot z$ that gives rise to (\ref{eq-mo}).
Applying Dirac's Hamiltonian formulation of the constrained systems to the latter Lagrangian, one finds the quadratic Hamiltonian \cite{ma-da,smilga-plb}:
    \be
    H=\frac{1}{2}\left(p_x^2+2x p_z+\alpha x^2-\beta z^2\right),
    \label{H1}
    \ee
that can be decoupled via a linear canonical transformations. A particular example is the transformation $(x,p_x,z,p_z)\to(x_1,p_1,x_2,p_2)$ defined by
    \bea
    x_1&:=&\frac{p_z+\omega^2_1 x}{\omega_1\sqrt{\omega_1^2-\omega_2^2}},~~~
    x_2:=\frac{p_x+\omega^2_1 z}{\sqrt{\omega_1^2-\omega_2^2}},
    \label{ct1}\\
    p_1&:=&\frac{\omega_1(p_x+\omega^2_2 z)}{\sqrt{\omega_1^2-\omega_2^2}},~~~
    p_2:=\frac{p_z+\omega^2_2 x}{\sqrt{\omega_1^2-\omega_2^2}},
    \label{ct2}
    \eea
where we have taken $\omega_1>\omega_2$ without loss of generality  \cite{smilga-plb}. This canonical transformation maps (\ref{H1}) to
    \be
    H={\frac{1}{2}\,(p_1^2+\omega_1^2x_1^2)}
    {\mathbf{-}\frac{1}{2}\,(p_2^2+\omega^2_2x_2^2)}.
    \label{H2}
    \ee

The standard canonical quantization of (\ref{H2}), $(x_1,p_1,x_2,p_2)\to
(\hat x_1,\hat p_1,\hat x_2,\hat p_2)$, with
    \be
    \begin{array}{ccc}
    (\hat x_j\psi)(x_1,x_2)&:=&x_j\psi(x_1,x_2),\\
    (\hat p_j\psi)(x_1,x_2)&:=&-i\frac{\partial}{\partial x_j}\psi(x_1,x_2),
    \end{array}
    \label{cq0}
     \ee
yields the Hermitian Hamiltonian operator,
    \be
    \hat H={\frac{1}{2}\,(\hat{p}_1^2+\omega_1^2\hat{x}_1^2)}
    {\mathbf{-}\frac{1}{2}\,(\hat{p}_2^2+\omega^2_2\hat{x}_2^2)}
    \label{q-H2}
    \ee
with eigenvalues: $\omega_1 (n_1+\frac{1}{2})-\omega_2 (n_2+\frac{1}{2})$. Here $\psi$ is an square-integrable function, $n_1,n_2=0,1,2,\cdots$, and we have used units in which $\hbar=1$. Because the spectrum of $\hat H$ is unbounded both from below and above, the corresponding quantum system is unstable \cite{footnote1}. One can avoid this problem by  performing an indefinite-metric quantization of the system, \cite{pauli-1943,sudarshan,nakanishi}. This yields a stable quantum theory which is, however, plagued with the presence of ghost states and the associated lack of unitarity. The same problem arises in the indefinite-metric quantization of higher-derivative theories of gravity \cite{stelle}.

In Ref.~\cite{be-ma}, the authors propose an alternative quantization of the classical Hamiltonian (\ref{H1}) that solves both the instability and non-unitarity problems associated with the standard definite- and indefinite-metric quantizations of the PU oscillator. In the following we first review the quantization scheme developed in \cite{be-ma} and show that it really amounts to the imaginary scaling, $x_2\to -i x_2$ and $p_2\to i p_2$, that clearly maps (\ref{H2}) into the standard classical Hamiltonian for a pair of decoupled simple harmonic oscillators. The latter can be easily quantized to yield a stable and unitary quantum system via the standard canonical quantization scheme. But it conflicts with the correspondence principle, as the classical limit of the resulting quantum theory does not coincide with the classical PU oscillator \cite{be-ma,pla-2010c}. This may be traced back to the fact that this quantization scheme involves mapping imaginary classical quantities to Hermitian operators. We show that these difficulties may be avoided, if we employ the quantum analog of the above imaginary scaling transformation. We explore the unusual properties of the quantum imaginary scaling transformation and use them to develop a general quantization scheme. This turns out to be a definitization of the indefinite-metric quantization that is related to the latter via a change of sign in the expression for the creation operator of the theory.
 \vspace{.3cm}

\noindent {\em {\bf 2. Imaginary-Scaling Quantization:}}~The main point of departure of the approach of Ref.~\cite{be-ma} is the imaginary scaling of dynamical variable $z$ of the Hamiltonian (\ref{H1}), namely $z\to y:=-iz$ and $p_z\to p_y:=ip_z$. This is a complex canonical transformation that maps (\ref{H1}) into a complex classical Hamiltonian whose standard canonical quantization, namely $(x,p_x,y,p_y)\to
(\hat x,\hat p_x,\hat y,\hat p_y)$ with
    \be
    \begin{array}{cc}
    (\hat x\psi)(x,y)=x\psi(x,y),&(\hat p_x\psi)(x,y)=-i\frac{\partial}{\partial x}\psi(x,y),\\
    (\hat y\psi)(x,y)=y\psi(x,y),&(\hat p_y\psi)(x,y)=-i\frac{\partial}{\partial y}\psi(x,y),
    \end{array}
    \label{cq2}
    \ee
yields the non-Hermitian $\cP\cT$-symmetric Hamiltonian \cite{be-ma}:
    \be
    \hat H_{_{\cP\cT}}:=\frac{1}{2}\left(\hat p_x^2-2 i \hat x \hat p_y+\alpha \hat x^2+\beta \hat y^2\right).
    \label{H-PT}
    \ee
It turns out that, for $\omega_1\neq\omega_2$, there is a similarity transformation \cite{jpa-2003,ph} that maps this operator to the sum of two simple harmonic oscillator Hamiltonians, $\hat h$. This shows that the spectrum of $\hat H_{_{\cP\cT}}$ is real and positive. Moreover, it implies the existence of a nonstandard inner product and the associated Hilbert space $\sH$ that restore the Hermiticity of $\hat H_{_{\cP\cT}}$, \cite{ph,ps-2010}. The quantum system \cite{ps-2010} defined by either of the unitary-equivalent Hilbert space-Hamiltonian pairs $(\sH,\hat H_{_{\cP\cT}})$  and $(L^2(\R^2),\hat h)$ is, therefore, both unitary and stable. The main difficulty with this scheme is that it yields a quantum theory that does not have the classical PU oscillator as its classical limit \cite{pla-2010c}. We will refer to this problem as the lack of a proper correspondence principle \cite{footnote0}.

Next, we recall that the similarity transformations mapping $\hat H_{_{\cP\cT}}$ to $\hat h$ correspond to (complex) linear canonical transformations. Because these commute with the canonical quantization of the underlying classical Hamiltonian, one may perform the necessary linear canonical transformations in the classical level and then quantize the system. This is particularly desirable, because it avoids dealing with the non-Hermitian Hamiltonian operator (\ref{H-PT}) and gives the same Hermitian Hamiltonian operator $\hat h$. Following this approach we first apply the real linear canonical transformation~(\ref{ct1})-(\ref{ct2}) on
the classical Hamiltonian (\ref{H1}) to obtain (\ref{H2}). We then perform the imaginary scaling transformation:
    \be
    (x_2,p_2)\to(-ix_2,ip_2)=:(\tilde x_2,\tilde p_2),
    \label{im-scale}
    \ee
that flips the unwanted sign in (\ref{H2}). Finally we quantize the resulting classical Hamiltonian to obtain:
    \be
    \hat h:=\frac{1}{2}\left(\hat p_1^2+\omega_1^2\hat x_1^2+
    \hat{\tilde p}_2^2+\omega_2^2\hat{\tilde x}_2^2\right).
    \ee
Clearly, this defines a unitary and stable quantum system. The procedure outlined in \cite{be-ma} can therefore be reduced to a simple imaginary scaling transformation. Note however that this simplification of the analysis of \cite{be-ma} does not provide a resolution of the difficulty related with the lack of a proper correspondence principle. An alternative root that avoids this difficulty is to perform the imaginary-scaling transformation on the quantum Hamiltonian operator (\ref{q-H2}). This is affected by a linear operator with rather peculiar properties.

Let $\sC^\infty$ denote the vector space of smooth complex-valued functions defined on the real line and $\hat x,\hat p$ be the standard position and momentum operators acting in $\sC^\infty$; for all $f\in\sC^\infty$,
    \be
    (\hat x f)(x):=x f(x),~~~~(\hat p f)(x):=-if'(x).
    \label{cq}
    \ee
Consider the operator $A:\sC^\infty\to\sC^\infty$ that is defined by
    \be
    A:=\exp\left(\frac{\pi}{4}\{\hat x,\hat p\}\right).
    \label{iso}
    \ee
It is not difficult to show that \cite{jpa-98}
    \be
    (Af)(x)=f(-ix).
     \ee
We can view $A$ as a linear operator acting in the Hilbert space of square-integrable functions $L^2(\R)$. This is a densely-defined unbounded one-to-one linear operator $A:L^2(\R)\to L^2(\R)$ with a dense range \cite{footnote2} that realizes the quantum imaginary scaling transformation:
    \be
    \begin{array}{c}
    \hat x\stackrel{A}{\longrightarrow}\hat{\tilde x}:=A\,\hat x\, A^{-1}=-i\hat x,\\
    \hat p\stackrel{A}{\longrightarrow}\hat{\tilde p}:=A\,\hat p\, A^{-1}=i\hat p.\end{array}
    \label{A-inv}
    \ee

Expression (\ref{iso}) suggests that $A$ is a Hermitian operator acting in $L^2(\R)$. This is however not true. One way of seeing this is to notice that if $A$ was a Hermitian operator, its square $A^2$, that is also densely defined, would have been a positive operator \cite{private}. This contradicts the fact that $g(x):=x e^{-x^4}$ is an eigenfunction of $A^2$ with eigenvalue $-1$, \cite{footnote11}. Therefore, $A$ is not Hermitian.

Another unusual but obvious property of $A$ is that the harmonic oscillator eigenfunctions do not belong to its domain, unless we consider it as mapping $L^2(\R)$ into another function space. Let $\psi_n(x):=N_n H_n(x) e^{-x^2/2}$ be the normalized eigenfunctions of the simple harmonic oscillator Hamiltonian $\hat\cH:=\frac{1}{2}({\hat p}^2+{\hat x}^2)$, with $N_n$ and $H_n$ being the normalization constants and Hermit polynomials, respectively. Clearly, $\psi_n\in\sC^\infty$ and
    \be
    \tilde\psi_n(x):=(A\psi_n)(x)=N_n H_n(-ix) e^{x^2/2}\notin L^2(\R).
    \ee
This observation suggests the definition of an appropriate Hilbert space $\tilde\sH$ containing $\tilde\psi_n$ and viewing $A$ as a linear operator mapping $L^2(\R)$ to $\tilde\sH$. In order to construct $\tilde\sH$, we endow the linear span $\tilde\sS$ of $\tilde\psi_n$ with the inner product:
    \be
    \bbr\tilde\phi,\tilde\psi\kkt:=\br A^{-1}\tilde\phi|A^{-1}\tilde\psi\kt=
    \int_{-\infty}^\infty \tilde\phi(ix)^*\tilde\psi(ix)dx,
    \label{inn-pro}
    \ee
and identify $\tilde\sH$ with the Cauchy completion of the resulting inner-product space \cite{reed-simon,footnote}.

By construction, for all $\phi,\psi\in L^2(\R)$ satisfying $A\phi,A\psi\in\tilde\sH$, we have $\br\phi|\psi\kt=\bbr A\phi,A\psi\kkt$. Therefore,
    \be
    A:L^2(\R)\to\tilde\sH
    \label{A}
    \ee
is a densely-defined isometry \cite{reed-simon}. In particular, it is a bounded operator with a bounded inverse whose domain includes $\psi_n$. As we will see momentarily this is a necessary step in relating the above imaginary-scaling quantization of the PU oscillator to its indefinite-metric quantization.

Now, consider the operator
    \be
    \hat{\tilde\cH}:=-\hat\cH=-\frac{1}{2}({\hat p}^2+{\hat x}^2).
    \label{tilde-cH}
     \ee
If we view $\hat{\tilde\cH}$ as an operator acting in $L^2(\R)$, it is a Hermitian operator with a negative spectrum given by $\{-n-\frac{1}{2}|n=0,1,2,\cdots\}$. A less obvious fact is that if we view $\hat{\tilde\cH}$ as an operator acting in $\tilde\sH$, it is again a Hermitian operator but its spectrum is positive. We can see this by performing the imaginary scaling transformation on $\hat{\tilde\cH}$, namely
    \be
    \hat{\tilde\cH}\to A^{-1}\hat{\tilde\cH}A=\frac{1}{2}({\hat p}^2+{\hat x}^2)=\hat\cH.
    \label{H5}
    \ee
The resulting operator, $\hat\cH$, acts in $L^2(\R)$ and has a positive spectrum, namely $\{n+\frac{1}{2}|n=0,1,2,\cdots\}$. Because $A:L^2(\R)\to\tilde\sH$ is a bounded operator with a bounded inverse and domain of $A$ contains the eigenfunctions $\psi_n$ of $\hat\cH $, the operators $\hat{\tilde\cH}:\tilde\sH\to\tilde\sH$  and  $\hat\cH :L^2(\R)\to L^2(\R)$ are isospectral. This shows that the spectrum of $\hat{\tilde\cH}$ is also real and positive. It is easy to see that $\tilde\psi_n$ are the eigenfunctions of $\hat{\tilde\cH}$ with eigenvalue $n+\frac{1}{2}$.

Let us recall that a quantum system is described by a Hilbert space and a Hamiltonian operator. The former determines the states (rays in the Hilbert space) and observables of the system (Hermitian operators acting in this Hilbert space), and the latter defines its dynamics. Two Hilbert space-Hamiltonian pairs $(\sH_1,\cH_1)$ and $(\sH_2,\cH_2)$ represent the same physical system if they are unitary-equivalent \cite{ps-2010}. This means that there is a unitary (inner-product preserving) operator $U:\sH_1\to\sH_2$ such that $\cH_2=U\cH_1U^{-1}$. It is easy to see that the pairs $(L^2(\R),\hat{\tilde \cH})$  and $(\tilde\sH,\hat{\tilde \cH})$ are not unitary-equivalent. Therefore they represent different quantum systems. We can obtain these by performing different quantizations of the same classical system, namely the one associated with the wrong-sign harmonic oscillator Hamiltonian:
    \be
    \tilde\cH:=-\frac{1}{2}(p^2+x^2).
    \label{classical}
    \ee
We obtain the quantum system described by $(L^2(\R),\hat{\tilde \cH})$, if we perform the standard canonical quantization of (\ref{classical}), namely
$x\to\hat x$ and $p\to\hat p$, where $\hat x$ and $\hat p$ act in $L^2(\R)$ and have the form (\ref{cq}). We obtain the quantum system described by $(\tilde\sH,\hat{\tilde \cH})$, if we perform the imaginary-scaling quantization of (\ref{classical}):
    \be
    x\to\hat{\tilde x},~~~~p\to\hat{\tilde p},
    \label{cq-3}
    \ee
where $\hat{\tilde x}$ and $\hat{\tilde p}$ are the operators of Eqs.~ (\ref{A-inv}) that act in $\tilde\sH$. Indeed, according to (\ref{tilde-cH}) and (\ref{A-inv}), $\hat{\tilde\cH}=\frac{1}{2}({\hat{\tilde p}}^2+{\hat{\tilde x}}^2)$, \cite{footnote12}.

It is important to notice that the imaginary-scaling quantization scheme involves a quantum-to-classical correspondence principle that associates to each (real) classical observable a Hermitian operator acting in the Hilbert space $\tilde\sH$.
It is this correspondence principle that gives physical meaning to quantum observables. For example,  (\ref{cq-3}) implies that the operators $\hat{\tilde x}=-i\hat x$ and $\hat{\tilde p}=i\hat p$ are the appropriate position and momentum observables of the quantum system $(\tilde\sH,\hat{\tilde \cH})$. Note that the standard position and momentum operators, $\hat x$ and $\hat p$, do not act as Hermitian operators in $\tilde\sH$. Therefore, they can no longer represent physical observables of the quantum system given by $(\tilde\sH,\hat{\tilde \cH})$.

Having established the basic features of the imaginary-scaling quantization and the ensuing correspondence principle, we can employ it for the quantization of the non-degenerate PU oscillator. We do this by expressing the classical Hamiltonian (\ref{H2}) as $H=H_1+H_2$, where $H_i=(-1)^{i+1}\frac{1}{2}(p_i^2+\omega_i^2x_i^2)$ with $i=1,2$. We then perform the standard canonical and imaginary-scaling quantizations of $H_1$ and $H_2$ on the Hilbert spaces $L^2(\R)$ and $\tilde\sH$, respectively. This gives rise to a quantum system with a Hilbert space $\check{\sH}$ and the Hamiltonian operator $\check{H}=\hat H_1+\hat{\tilde H}_2$ that describes a pair of uncoupled simple harmonic oscillators. The Hilbert space $\check{\sH}$ consists of functions $\psi:\R^2\to\C$ fulfilling $\int_{-\infty}^\infty dx_1\int_{-\infty}^\infty dx_2|\psi(x_1,ix_2)|^2<\infty$; its inner product has the form
    \be
    \bbr\phi|\psi\kkt:=\int_{-\infty}^\infty dx_1\int_{-\infty}^\infty dx_2\,     \phi(x_1,ix_2)^*\psi(x_1,ix_2);
    \ee
and the quantization scheme involves
    \be
    x_1\to\hat x_1,~~~p_2\to\hat p_2,~~~
    x_2\to\hat{\tilde x}_2:=-i\hat x_2,~~~p_2\to\hat{\tilde p}_2:=i\hat p_2,\nn
    \ee
 where $\hat x_1,\hat p_1$ and $\hat{\tilde x}_2,\hat{\tilde p}_1$ act in $\check\sH$ according to (\ref{cq0}) and
    \be
    \begin{array}{ccc}
    (\hat{\tilde x}_2\psi)(x_1,x_2)&:=&-ix_2\psi(x_1,x_2),\\
    (\hat{\tilde p}_2\psi)(x_1,x_2)&:=&\frac{\partial}{\partial x_2}\psi(x_1,x_2),
    \end{array}
    \nn
    \ee
respectively.

Note that unlike the approach of Ref.~\cite{be-ma}, our proposal for quantizing the non-degenerate PU oscillator does not involve complex classical quantities (or real classical quantities with imaginary time derivatives.) The classical PU oscillator is described by the phase space $\R^4$, that is parameterized by the classical position and momentum variables $(x_1,x_2,p_1,p_2)$, and the classical Hamiltonian $H$ that generates its dynamics. The imaginary-scaling quantization of this system yields a quantum PU oscillator that is represented by the Hilbert space $\check{\sH}$  and the Hamiltonian operator $\check{H}$. The dynamics of the system is determined by the latter through the time-dependent Schr\"odinger equation, $i\dot\psi(t)=\check{H}\psi(t)$ with $\psi(t)\in\check\sH$, in the Schr\"odinger picture. Alternatively, we can employ the Heisenberg equation,
    \be
    i\frac{d}{dt}\,{\check A}(t)=[\check A(t),\check H],
    \label{HE}
    \ee
for observables $\check A(t)$ in the Heisenberg picture. In particular, the Heisenberg-picture position and momentum operators $\hat x_1(t)$, $\hat{\tilde x}_2(t)$, $\hat p_1(t)$, $\hat{\tilde p}_2(t)$ that are defined by
    \bea
    &&\hat x_1(t):=e^{it\check H}\hat x_1 e^{-it\check H},~~~~\hat{\tilde x}_2(t):=e^{it\check H}\hat{\tilde x}_2 e^{-it\check H},
    \label{po-1}\\
    &&\hat p_1(t):=e^{it\check H}\hat p_1 e^{-it\check H},~~~~\hat{\tilde p}_2(t):=e^{it\check H}\hat{\tilde p}_2 e^{-it\check H},~~~~~~
    \label{mo-1}
    \eea
are solutions of (\ref{HE}). It is indeed a simple exercise to show that the Heisenberg equation (\ref{HE}) for the position and momentum operators take the form
    \bea
    &&\frac{d}{dt}\hat x_1(t)=\hat p_1(t),~~~~
    \frac{d}{dt}\hat p_1(t)=-\omega_1^2\hat p_1(t),\nn\\
    &&\frac{d}{dt}\hat{\tilde x}_2(t)=\hat{\tilde p}_2(t),~~~~
    \frac{d}{dt}\hat{\tilde p}_2(t)=-\omega_2^2\hat{\tilde x}_2(t).\nn
    \eea
Hence the position operators $\hat x_1(t)$ and $\hat{\tilde x}_2(t)$
satisfy the classical equations of motion: $\ddot x_1(t)+\omega_1^2x_1(t)=0$ and $\ddot x_2(t)+\omega_2^2x_2(t)=0$, respectively.

The main subtlety of our imaginary-scaling quantization scheme is that it maps the classical Hamiltonian $H$, which as a real-valued function of the phase space is not bounded from below, to the quantum Hamiltonian $\check H$ that possesses a positive spectrum. This does not cause a difficulty, because $\check H$ enters into our formalism as the generator of time-evolution. If we interpret the classical Hamiltonian (that generates the time-evolution) also as an energy observable with unbounded negative and positive values, then clearly the quantum Hamiltonian cannot be the corresponding quantum energy observable \cite{footnote21}. If we insist on identifying the quantum Hamiltonian with a quantum energy observable, then the correspondence principle we use differs from the standard correspondence principle, because ours relates a classical energy observable with negative as well as positive values to a quantum energy observable that has a positive spectrum. Note that any attempt to quantize the PU oscillator that yields a Hermitian Hamiltonian with a bounded-below spectrum should necessarily involve a correspondence principle similar to ours \cite{footnote4}.
\vspace{.3cm}

\noindent {\bf 3.~Connection to Indefinite-Metric Quantization:} Consider the standard annihilation and creation operators, $\hat a:=(\hat x+i\hat p)/\sqrt 2$ and $\hat a^\dagger:=(\hat x-i\hat p)/\sqrt 2$, that satisfy  $[\hat a,\hat a^\dagger]=1$, $\hat a\psi_0=0$, and $\psi_n=(n!)^{-1/2}{\hat a}^{\dagger n}\psi_0$. We can view $\psi_n$ as elements of $\sC^\infty$ and try to reconstruct the Hilbert space of the state vectors of the simple Harmonic oscillator $\hat\cH $, namely $L^2(\R)$, using $\psi_n$. This is done by endowing the linear span $\sS$ of  $\psi_n$ with an inner product $\br\cdot,\cdot\kt$ that renders the number operator $\hat N:=\hat a^\dagger\hat a$ Hermitian. This holds if $\hat a^\dagger$ is the adjoint of $\hat a$, i.e., for all $\phi,\psi\in\sS$, $\br\phi,a\psi\kt=\br a^\dagger\phi,\psi\kt$. Once we select an inner product fulfilling this condition, we can identify the Hilbert space with the Cauchy completion of $\sS$. If we use the $L^2$-inner product, $\br\psi|\phi\kt=\int_{-\infty}^\infty\phi(x)^*\psi(x)dx$, that has the above property, we obtain $L^2(\R)$ as the Hilbert space of the system.

Next, following the standard argument used in the indefinite-metric quantization scheme, we propose to construct a Hilbert space using a different set of functions, ${\tilde\phi}_n$, that we define as follows: $\tilde\phi_0$ is (up to the choice of a normalization constant) a solution of the differential equation $\hat a^\dagger\tilde\phi_0=0$, and ${\tilde\phi}_n:=(n!)^{-1/2}{\hat a}^n\tilde\phi_0$  for all $n=1,2,\cdots$. In other words, we now interpret $\hat{\tilde a}:=\hat a^\dagger$ and $\hat{\tilde a}^\dagger:=\hat a$ as the annihilation and creation operators, respectively. This is because
    \be
    \hat{\tilde a}\,\tilde\phi_0=0,~~~~{\tilde\phi}_n=(n!)^{-1/2}{\hat{\tilde a}}^{\dagger n}\tilde\phi_0.
    \label{tilde}
    \ee
The new annihilation and creation operators, $\hat{\tilde a}$ and $\hat{\tilde a}^\dagger$ satisfy the ``abnormal'' bosonic commutation relation  \cite{sudarshan}:
    \be
    [\hat{\tilde a},\hat{\tilde a}^\dagger]=-1.
    \label{abnormal}
    \ee

We will denote the linear span of $\tilde\phi_n$ with $\tilde\sT$ and try to construct an appropriate inner product $\pbr\cdot,\cdot\pkt$ on $\tilde\sT$ that makes $\hat{\tilde a}^\dagger$ adjoint of $\hat{\tilde a}$, i.e., for all $\tilde\xi,\tilde\zeta\in\tilde\sT$,
    \be
    \pbr\tilde\xi,\hat{\tilde a}\,\tilde\zeta\pkt=
    \pbr \hat{\tilde a}^\dagger\tilde\xi,\tilde\zeta\pkt.
    \label{inn-condi}
    \ee
It turns out that the number operator for $\tilde\psi_n$ is the operator
$\hat{\tilde N}:\tilde\sT\to\tilde\sT$ that is defined by
    \be
    \hat{\tilde N}:=-\hat{\tilde a}^\dagger\hat{\tilde a}=-\hat a\hat a^\dagger.
    \label{tilde-N}
    \ee
We can use (\ref{tilde}) and (\ref{abnormal}) to show that $\hat{\tilde N}\tilde\phi_n=n\tilde\phi_n$.  Furthermore, in light of (\ref{tilde-N}), we have $\hat{\tilde\cH}=\hat{\tilde N}+\frac{1}{2}$. This shows that if we can find an inner product respecting (\ref{inn-condi}), we can complete $\tilde\sT$ into a Hilbert space $\sH'$ and view $\hat{\tilde\cH}$ as an operator acting in $\sH'$. The spectrum of this operator will then consist of the eigenvalues $n+\frac{1}{2}$ with $n=0,1,2,\cdots$. In particular, it will be positive.

The main difficulty with the above construction is that the condition~(\ref{inn-condi}) conflicts with the positive-definiteness of the inner product $\pbr\cdot,\cdot\pkt$; one can use (\ref{tilde}) and (\ref{abnormal}) to show that for all $m,n=0,1,2,\cdots$,
    \be
    \pbr\tilde\phi_m,\tilde\phi_n\pkt=(-1)^n\delta_{mn}\pbr
    \tilde\phi_0,\tilde\phi_0\pkt.
    \label{indef-1}
    \ee
Therefore, $\pbr\cdot,\cdot\pkt$ is an indefinite inner product. It gives $\tilde\sT$ the structure of a Krein space. It is customary to choose $\pbr\tilde\phi_0,\tilde\phi_0\pkt=1$ and view $\tilde\phi_{2n+1}$, that have an imaginary norm, as defining ``ghost states.''

A key observation that links the imaginary-scaling and indefinite-metric quantization schemes is that up to a constant coefficient $\tilde\phi_n$ coincides with $\tilde\psi_n$; as elements of $\sC^\infty$, they are related according to
    \be
    \tilde\phi_n=i^n \tilde\psi_n.
    \label{link}
    \ee
This in turn implies that $\tilde\phi_n$ belong to the Hilbert space $\tilde\sH$ of the preceding section, and $\tilde\sT=\tilde\sS$.

The relationship between the indefinite inner product $\pbr\cdot,\cdot\pkt$ and the definite inner product $\bbr\cdot,\cdot\kkt$ is typical of the Hilbert spaces endowed with a Krein-space structure \cite{cjp}. To describe this relationship we use the parity operator $\cP$ to introduce
    \be
    \Pi_\pm:=\frac{1}{2}(1\pm\cP),~~~~\tilde\sS_\pm:=\Pi_\pm\tilde\sS.
     \label{ops}
     \ee
It is easy to see that $\tilde\psi_{2n}\in\tilde\sS_+$, $\tilde\psi_{2n+1}\in\tilde\sS_-$,
    \be
    \tilde\sS=\tilde\sS_+\oplus\tilde\sS_-,
    \label{dec}
    \ee
and $\tilde\sS_\pm$ are the eigenspaces of the restriction of $\cP$ onto $\tilde\sS$ with eigenvalues $\pm1$. Note also that in view of (\ref{inn-pro}), (\ref{indef-1}), and (\ref{link}), this is an orthogonal decomposition with orthogonality condition defined by either of $\bbr\cdot,\cdot\kkt$ and $\pbr\cdot,\cdot\pkt$. Furthermore, we can use  (\ref{inn-pro}), (\ref{indef-1}), and (\ref{link}) to show that for all $\tilde\xi,\tilde\zeta\in\tilde\sS$,
    \bea
    \bbr\tilde\xi,\tilde\zeta\kkt&=&
    \pbr\Pi_+\tilde\xi,\Pi_+\tilde\zeta\pkt-
    \pbr\Pi_-\tilde\xi,\Pi_-\tilde\zeta\pkt,
    \label{link2}\\
    \pbr\tilde\xi,\tilde\zeta\pkt&=&\bbr\tilde\xi,\cP\tilde\zeta\kkt=
    \int_{-\infty}^\infty\tilde\xi(ix)^*\tilde\zeta(-ix)dx.
    \label{link3}
    \eea
To the best of our knowledge, the explicit form of the indefinite-metric inner product given by Eq.~(\ref{link3}) was not previously reported in the extensive literature on the subject \cite{nakanishi}.

\noindent {\bf 4.~Creation and Annihilation Operators:} The above analysis suggests that the number operator $\tilde N$ has a real spectrum and an orthonormal set of eigenvectors belonging to the Hilbert space $\tilde\sH$, namely $\tilde\psi_n$. This shows that $\tilde N$ is a Hermitian operator acting in $\tilde\sH$. However, we also know that every inner product that identifies $\hat{\tilde a}^\dagger$ with the adjoint of $\hat{\tilde a}$ is necessarily indefinite. Therefore, as operators acting in $\tilde\sH$, $\hat{\tilde a}^\dagger$ is not the adjoint of $\hat{\tilde a}$, yet $\tilde N=-\hat{\tilde a}^\dagger\hat{\tilde a}$ is a Hermitian (self-adjoint) operator. In order to see that this strange phenomenon does not lead to any inconsistency, we make the following observations. First, we note that according to (\ref{inn-pro}) and (\ref{link2}),
    \be
    \pbr\tilde\xi,\cP\tilde\zeta\pkt=\pbr\cP\tilde\xi,\tilde\zeta\pkt.
    \label{eq1}
    \ee
Combining this relation with (\ref{link2}) and using the fact that $\cP$ and $\hat{\tilde a}^\dagger$ anticommmute, we have
    \be
    \bbr\tilde\xi,\hat{\tilde a}\tilde\zeta\kkt=
    \bbr\cP\hat{\tilde a}^\dagger\cP\tilde\xi,\tilde\zeta\kkt=
    \bbr-\hat{\tilde a}^\dagger\tilde\xi,\tilde\zeta\kkt.
    \ee
This identifies the adjoint of $\hat{\tilde a}$ with $-\hat{\tilde a}^\dagger$. Denoting the latter by $\tad$, we find that $\tilde N=-\hat{\tilde a}^\dagger\hat{\tilde a}=\tad\hat{\tilde a}$. Therefore $\tilde N$ is indeed a positive (and in particular Hermitian) operator acting in $\tilde\sH$. Another outcome of this calculation is that we can reconstruct the Hilbert space of the imaginary-scaling quantization by, respectively, adopting
    \be
    \hat{\tilde a}:=\hat a^\dagger,~~~~\tad:=-\hat a,
    \ee
as the annihilation and creation operators. Because these satisfy the standard bosonic commutation relation: $[\hat{\tilde a},\tad]=1$, the algebra of the creation and annihilation operators of the imaginary-scaling quantization is identical with the usual canonical quantization of a bosonic system. What makes the difference is how the creation and annihilation operators are related to the standard position and momentum operators (quantized fields and their time-derivative in field theory). \vspace{3mm}

\noindent {\bf 6.~Concluding Remarks:}~ In this paper we have established the imaginary scaling transformation as the main ingredient of the quantization scheme developed in Ref.~\cite{be-ma}. We have subsequently simplified this scheme in such a way that it avoids dealing with non-Hermitian $\cP\cT$-symmetric operators. We have examined the consequences of performing the quantum imaginary scaling transformation. This is realized by a linear operator with rather unusual properties. We have exploited these properties to identify the appropriate Hilbert space of state vectors of the non-degenerate quantum PU oscillator. We have outlined a quantization scheme involving a classical correspondence principle that differs from the standard canonical quantization in the definition of the Hilbert space of state vectors. As in the standard quantum mechanics, the choice of the Hilbert space encodes the boundary conditions of the problem. Different choices lead to different physical systems. What we have done is to outline a quantization scheme involving a particular choice for the Hilbert space that maps the classical non-degenerate PU oscillator to a particular unitary and stable quantum system. Similarly to the approach of Ref.~\cite{be-ma}, it involves the imaginary-scaling transformation which is however employed in the quantum level. Its main advantage over the proposal of Ref.~\cite{be-ma} is that it avoids using complex classical quantities, mapping them into Hermitian operators upon quantization, or producing a quantum system that does not seem to yield classical PU oscillator in the classical limit.

We have examined the relationship between the imaginary-scaling and indefinite-metric quantization schemes and shown that they share the same vector space of state vectors. We have derived, for the first time, an explicit expression for the standard indefinite inner product and demonstrated that it gives the Hilbert space of the imaginary-scaling quantization the structure of a Krein space.

We have obtained an appropriate pair of creation and annihilation operators for the imaginary-scaling quantization scheme. These differ from the creation and annihilation operators of the indefinite-metric theories in the sign of the creation operator. This seemingly minor difference is responsible for transforming the ghosts into physical states and the subsequent restoration of unitarity. It is important to note that this is achieved at no energy cost; the contribution of the ghost states to the spectrum of the Hamiltonian is the same as in the corresponding indefinite-metric theory.  For instance, if we consider the sum of two harmonic oscillator Hamiltonians with unit mass and frequency, and quantize the first of these using the standard canonical quantization and the second via the imaginary scaling quantization, the energy spectrum turns out to be the difference of the mode numbers. In particular, similarly to the case of indefinite-metric quantization of the second oscillator, the vacuum energy vanishes identically \cite{pauli-1943}. This observation seems to indicate that the imaginary-scaling quantization shares the niceties of the indefinite-metric quantization while not suffering from the lack of a consistent probabilistic interpretation. We plan to examine the prospects of this scheme in dealing with the quantum mechanical and field theoretical models that were treated in the context of indefinite-metric quantum theories.

As a final comment we wish to stress that although a direct extension of our results for PU oscillators involving 6 or higher (even) order derivatives is straightforward, the same cannot be said about higher-order derivative theories that include nonlinear interactions. The latter may lead to unsurmountable technical as well as conceptual difficulties, particularly in connection with the correspondence principle and the energy observable. Therefore, one should exercise great care in any attempt to use our formalism for such theories \cite{footnote31}.\vspace{.3cm}

\noindent {\bf Acknowledgments:} I wish to express my gratitude to Uwe G\"unther, Sergii Kuzhel, Boris Samsonov, Bayram Tekin, and Miloslav Znojil for helpful discussions. This work has been supported by the Turkish Academy of Sciences (T\"UBA).

\ed
\begin{thebibliography}{99}
\bibitem{footnote1} As pointed out in Refs.~\cite{robert-smilga,smilga-sigma} because PU-oscillator does not involve nonlinear interactions, strictly speaking the fact that the spectrum of the Hamiltonian is not bounded below does not cause any serious problem. However the inclusion of generic nonlinear interactions results in the emergence of an instability.
\bibitem{footnote0} Throughout this article we use the term ``correspondence principle'' to mean a clear rule that assigns to each operator describing an observable of a quantum system a particular classical observable that is by definiton a real-valued function of the phase space of the undelying classical system. It is this correspondence that gives physical meaning to the quantum observables.
\bibitem{footnote2} Smooth functions having a compact support belong to the domain of $A$. Because these functions constitute a dense subset of $L^2(\R)$, $A$ is densely defined. It is also easy to see that $A^{-1}={\cal P}A$. Therefore, as operators acting in $L^2(\R)$, the domains of $A$ and $A^{-1}$ coincide.
\bibitem{footnote11} $A^2$ is the restriction of the parity operator $\cP$ on the domain of $A^2$. The parity operator is defined by $(\cP\psi)(x):=\psi(-x)$ for every function $\psi:\R\to\C$ defined on $\R$.
\bibitem{footnote} It is tempting to view equation~(\ref{inn-pro}) as an indication that $\bbr\tilde\xi,\tilde\zeta\kkt$ can be expressed as $\br\tilde\xi|\eta_{_+}\tilde\zeta\kt$ with $\eta_{_+}$ being the metric operator $\eta_{_+}=A^{-1\dagger}A^{-1}$. This argument is erroneous, because in general $\tilde\xi$ and $\eta_{_+}\tilde\zeta$ do not belong to $L^2(\R)$, and $\br\tilde\xi|\eta_{_+}\tilde\zeta\kt$ may be divergent.
\bibitem{footnote12} We can use (\ref{inn-pro}) and (\ref{H5}) to show that the pairs $(\tilde\sH,\hat{\tilde \cH})$ and $(L^2(\R),\hat{\cH})$ are unitary-equivalent \cite{ph}. As a result, they describe the same quantum system, namely a simple harmonic oscillator.
\bibitem{footnote21} One may actually question whether the classical Hamiltonian $H$ of the PU oscillator can be identified with an energy observable.
\bibitem{footnote4} To apply our approach for quantizing higher-derivative physical theories one needs to address the issue of the energy operator. Our point of view is that our apprach maps classical real variables to Hermitian operators acting in a Hilbert space and that the dynamics is generated by a Hermitian operator with a spectrum that is bounded from below. In this sense the quantum system is both unitary and stable. Imposing additional conditions such as identifying the Hamiltonian with an energy operator and requiring that it should correspond to a classical energy observable lacking a lower bound will definitely rule out our scheme as an acceptable quantization of the given classical system. Note however that this condition also rules out any other quantization scheme that yields a stable quantum system.
\end{thebibliography}
